\documentclass[a4paper]{article}
\usepackage{amsmath}
\usepackage{graphicx}

\begin{document}

\title{Gaussian relative entropy of entanglement }
\author{Xiao-yu Chen \\
School of Science, China Institute of Metrology, 310018, Hangzhou, China}
\date{}
\maketitle

\begin{abstract}
For two gaussian states with given correlation matrices, in order that
relative entropy between them is practically calculable, I in this paper
describe the ways of transforming the correlation matrix to matrix in the
exponential density operator. Gaussian relative entropy of entanglement is
proposed as the minimal relative entropy of the gaussian state with respect
to separable gaussian state set. I prove that gaussian relative entropy of
entanglement achieves when the separable gaussian state is at the border of
separable gaussian state set and inseparable gaussian state set. For two
mode gaussian states, the calculation of gaussian relative entropy of
entanglement is greatly simplified from searching for a matrix with 10
undetermined parameters to 3 variables. The two mode gaussian states are
classified as four types, numerical evidence strongly suggests that gaussian
relative entropy of entanglement for each type is realized by the separable
state within the same type.For symmetric gaussian state it is strictly
proved that it is achieved by symmetric gaussian state.

PACS number:03.67.Mn, 03.65.Ud
\end{abstract}

\section{Introduction}

Quantum relative entropy function has many applications in the problems of
classical and quantum information transfer and quantum data compression \cite
{Schumacher}. The relative entropy has a natural interpretation in terms of
the statistical distinguishability of quantum states; closely related to
this is the picture of relative entropy as a distance measure between
density operators. Based on the relative entropy, a nature measure of
entanglement called the relative entropy of entanglement was proposed. This
entanglement measure is intimately related to the entanglement of
distillation by providing an upper bound for it. \ It tells us that the
amount of entanglement in the state with its distance from the disentangled
set of states. In statistical terms, the more entangled a state is the more
it is distinguishable from a disentangled state\cite{Vedral}. However,
except for some special situations\cite{Wu}, such an entanglement measure is
usually very difficult to be calculated for mixed state. For continuous
variable system, it is shown that the relative entropy of entanglement is
actually trace-norm continuous and hence well-defined even in this
infinite-dimensional context\cite{Eisert}. Due to the fact that gaussian
state has the merit that the logarithmic of the state operator is in the
quadrature form of canonical operators, the distance of the gaussian state
to gaussian state measured by the relative entropy (gaussian relative
entropy or GRE) was considered \cite{Scheel}. In which the second gaussian
state was specified by its exponential operator matrix (EM), not by the
usual correlation matrix (CM). Until now, an explicit transform from the CM
to EM has not been available, make the calculation of the relative entropy
between gaussian states in fact impossible. Moreover, when using the
gaussian relative entropy as entanglement measure, the second state should
be separable. Now all the separable criterions are not in the form of EM. So
the problems of separable criterion in the form of EM or the explicit
transform of CM to EM need to be addressed. In this paper, I will give the
explicit transform of CM to EM, propose the gaussian relative entropy of
entanglement (GREE) and give the method of how to calculate it. The paper is
organized as follow: In section 2, the transform of CM to EM for $q-p$
decorrelation gaussian state is given with the help of symplectic
transformation, further more, the direct transforms of any CM to EM and EM
to CM are also solved due to the commutation relation of matrices. Section 3
deals with GREE with emphasis on the proof of theorem 2 which states that
GREE will achieve by gaussian state at border of separable and inseparable
gaussian state sets. In section 4, I concentrate on the simplification of
GREE for two mode ($1\times 1$) gaussian state system. In section 5, the $%
1\times 1$ gaussian states are classified as four types, their GREEs are
discussed. In section 6, conclusions and discussions are addressed.

\section{Matrix in the exponential density operator}

To characterize a gaussian state, we have several equivalent means, among
them are: quantum characteristic function specified by first (irrelative to
the entanglement problem) and second moments which are also called means and
CM, density operator in exponential form specified by a matrix M
(exponential operator matrix or EM), density operator in exponential form of
ordered operators specified by another matrix (ordered exponential operator
matrix or OEM). The separability of a gaussian state was obtained with CM
\cite{Duan}\cite{Simon}\cite{Giedke}, also with OEM \cite{Wang}. The
transform of CM to OEM and vice versa are quite directly by the integral
within ordered product of operators. The transform of EM to OEM is also
available but involved with a calculation of exponential of matrix\cite
{Wang1}. Scheel \cite{Scheel} derived a relation of EM and CM of gaussian
state with generation and annihilation operators. Following the way I now
derive their relation with canonical operators.

Gaussian quantum state can be given by the density operator

\begin{equation}
\rho =\frac{\exp [-\frac{1}{2}F^{T}MF]}{Tr\{\exp [-\frac{1}{2}F^{T}MF]\}},
\label{wave1}
\end{equation}
where $M$ is a real symmetric matrix, and $F=\left[ q_{1},\cdots
q_{n};p_{1},\cdots p_{n}\right] ^{T}$ , with $q_{j},p_{j}$ are the canonical
operators. In order to relate the matrix $M$ to the correlation matrix (CM) $%
\alpha $, a unitary transformation

\begin{equation}
F^{^{\prime }}=UFU^{-1}=SF
\end{equation}
is needed. The matrix $S$ produces a symplectic transformation on the
canonical operators $F$. To preserve the commutation relations of the
canonical operators, $S$ should satisfy the symplectic condition $S\Delta
S^{T}=\Delta $. Where

\begin{equation}
\Delta =\left[
\begin{array}{cc}
0 & I_{n} \\
-I_{n} & 0
\end{array}
\right] .
\end{equation}
The matrix $S$ is chosen such that it diagonalizes $M$ , hence $S^{T}MS=$ $%
\widetilde{M}$ (with $\widetilde{M}$ being diagonal). Then the
characteristic function of the density operator (\ref{wave1}) is

\begin{eqnarray}
Tr[\rho \exp (iF^{T}z)] &=&Tr[U\rho U^{-1}U\exp (iF^{T}z)U] \\
&=&\exp [-\frac{1}{2}z^{T}S(\frac{1}{2}\coth \frac{1}{2}\widetilde{M}%
)S^{T}z],  \notag
\end{eqnarray}
and the CM is $\alpha =S(\frac{1}{2}\coth \frac{1}{2}\widetilde{M})S^{T}=S%
\widetilde{\alpha }S^{T}$. In the derivation the expression for
characteristic function of a thermal state has been used.

For a given CM, how to find such a sympleptic transformation is the topics
of this section. To find the $S$ matrix, let us consider $\Delta ^{-1}\alpha
$ instead of $\alpha $, then

\begin{equation}
S^{T}\Delta ^{-1}\alpha (S^{T})^{-1}=\Delta ^{-1}\widetilde{\alpha }.
\end{equation}
The eigenvalues of $\Delta ^{-1}\alpha $ come in pairs $\pm i\gamma _{j}$,
the matrix uncertainty relation requires $\gamma _{j}\geq \frac{1}{2}$ \cite
{Holevo}, where $\gamma _{j}$ are called symplectic eigenvalues of $\alpha $
\cite{Vidal}, then $\widetilde{\alpha }=diag\{\gamma _{1},\cdots ,\gamma
_{n},\gamma _{1},\cdots ,\gamma _{n}\}$, and $\widetilde{M}=diag\{\widetilde{%
M}_{1},\cdots ,\widetilde{M}_{n},$ $\widetilde{M}_{1},\cdots ,\widetilde{M}%
_{n}\}$, with $\widetilde{M}_{j}=\log \frac{2\gamma _{j}+1}{2\gamma _{j}-1}$
are symplectic eigenvalues of $M$. Let $\widetilde{\Psi }$ be the
eigenvector of $\Delta ^{-1}\widetilde{\alpha }$ and $\Psi $ be the
eigenvector of $\Delta ^{-1}\alpha $, then

\begin{equation}
\Psi =(S^{T})^{-1}\widetilde{\Psi }.  \label{wave2}
\end{equation}
So that $S$ matrix can obtained from the eigenvectors. The $S$ matrix so
obtained is not totally determined, yet the symplectic condition $S\Delta
S^{T}=\Delta $ should be verified.

If the CM is in its $q-p$ decorrelation form, that is $\alpha =\alpha
_{q}\oplus \alpha _{p}$, then the details of $S$ can be further worked out.
The $q-p$ decorrelation CMs are quite general in concerning with
entanglement for $1\times 1$ gaussian states. For these states, the CMs can
always be transformed into the form of $\alpha _{q}\oplus \alpha _{p}$ by
local operations\cite{Duan} \cite{Simon}. For multi-mode bipartite gaussian
state, It is not known if the CM can be transformed to the form of $\alpha
_{q}\oplus \alpha _{p}$ or not just by local operations. Let us start with
the CM of the form $\alpha _{q}\oplus \alpha _{p}$, that is to say the
correlation between position and momentum of each mode and inter-modes have
already dissolved, so that one just considers a symplectic transformation of
form $S=S_{q}\oplus S_{p}$. The relation $S\Delta S^{T}=\Delta $ then will
requires that $S_{q}S_{p}^{T}=S_{p}S_{q}^{T}=I$. And $S$ will take the form
of $S=S_{q}\oplus (S_{q}^{T})^{-1}$, and $(S^{T})^{-1}=(S_{q}^{T})^{-1}%
\oplus S_{q}$. $S^{-1}\alpha (S^{T})^{-1}=[S_{q}^{-1}\alpha
_{q}(S_{q}^{T})^{-1}]\oplus \lbrack S_{q}^{T}\alpha _{p}S_{q}]=\widetilde{%
\alpha }$. Let $\Psi ^{T}=(\Psi _{q}^{T},\Psi _{p}^{T})$ and $\widetilde{%
\Psi }^{T}=(\widetilde{\Psi }_{q}^{T},\widetilde{\Psi }_{p}^{T})$, then the
eigenequation $\Delta ^{-1}\alpha \Psi =\pm i\gamma \Psi $ (where $\gamma $
is one of the symplectic eigenvalues of $\alpha $) will be \ $-\alpha
_{p}\Psi _{p}=\pm i\gamma \Psi _{q}$ and $\alpha _{q}\Psi _{q}=\pm i\gamma
\Psi _{p}$. And one gets
\begin{equation}
\alpha _{p}\alpha _{q}\Psi _{q}=\gamma ^{2}\Psi _{q}\text{, \ \ \ }\alpha
_{q}\alpha _{p}\Psi _{p}=\gamma ^{2}\Psi _{p}.  \label{wave0}
\end{equation}
For the eigenvalue $i\gamma _{j}$, suppose the eigenvector of the $q$ part
be $\Psi _{qj}^{T}=[c_{j1,}\cdots ,c_{jn}].$ Since $\alpha _{p}\alpha
_{q}\Psi _{q}^{\ast }=\gamma ^{2}\Psi _{q}^{\ast },$ the eigenvector $\Psi
_{q}$ can always be chosen to be real. Then $\Psi _{pj}^{T}=-i[c_{j1,}\cdots
,c_{jn}]\alpha _{q}/\gamma _{j}$. The similar equation for \ $\widetilde{%
\Psi }$ will at last give the result of

\begin{equation}
\widetilde{\Psi }_{j}^{T}=\frac{1}{\sqrt{2}}\left[ 0,\cdots ,0,1,0,\cdots
,0;0,\cdots ,0,-i,0,\cdots ,0\right] \text{ \ \ for eigenvalue }i\gamma _{j},
\end{equation}
where the nonzero elements in the eigenvectors are at the positions of $j$
and $n+j$. By Eq.(\ref{wave2}) one gets $\widetilde{\Psi }^{T}(S^{T})^{-1}%
\widetilde{\Psi }=\widetilde{\Psi }^{T}\Psi $ and $\widetilde{\Psi }%
^{+}(S^{T})^{-1}\widetilde{\Psi }=\widetilde{\Psi }^{+}\Psi $, then

\begin{eqnarray}
(S_{q}^{T})_{kj}^{-1} &=&c_{jk}  \label{wave3} \\
(S_{q})_{kj} &=&\sum_{l}c_{jl}(\alpha _{q})_{lk}/\gamma _{j}.  \label{wave4}
\end{eqnarray}
One has to verify the consistence of Eq.(\ref{wave3}) and Eq.(\ref{wave4}),
that is $S_{q}^{-1}S_{q}=I_{n}$. The vector $\Psi _{qj}$ has yet a phase
factor and its length left to be determined. To determine the phase factor
of $\Psi _{qj},$ one simply chooses $c_{jj}$ to be positive. The length of
the vector $\Psi _{qj}$ can be chosen such that
\begin{equation}
\sum_{kl}c_{jk}c_{jl}(\alpha _{q})_{lk}/\gamma _{j}=1.
\end{equation}
So $\sum_{k}\left( S_{q}^{-1}\right) _{jk}\left( S_{q}\right) _{kj}=1.$
While in proving $\sum_{k}\left( S_{q}^{-1}\right) _{mk}\left( S_{q}\right)
_{kj}=0$ for $m\neq j$, one has $\Psi _{pj}^{+}(\alpha _{q}\alpha
_{p}-\alpha _{q}\alpha _{p})\Psi _{qm}=$ $\Psi _{pj}^{+}(\left( \alpha
_{p}\alpha _{q}\right) ^{+}-\alpha _{q}\alpha _{p})\Psi _{qm}=\left( \gamma
_{j}^{2}-\gamma _{m}^{2}\right) \Psi _{pj}^{+}\Psi _{qm}=0.$ Hence, for $%
\gamma _{m}\neq \gamma _{k}$, one gets $\Psi _{pj}^{+}\Psi _{qm}=0$, and the
proof of the consistence of Eq.(\ref{wave3}) and Eq.(\ref{wave4}) is
completed.

Hence the symplectic transformation can be constructed directly from the
eigenvectors of $\alpha _{p}\alpha _{q}$ or equivalently $\alpha _{q}\alpha
_{p}$. As an example, let us consider the 1$\times $1 symmetric gaussian
state. The $q$ and $p$ parts of CM are

\begin{equation}
\alpha _q=\frac 12\left[
\begin{array}{ll}
m & k_q \\
k_q & m
\end{array}
\right] ,\text{ \quad }\alpha _p=\frac 12\left[
\begin{array}{ll}
m & -k_p \\
-k_p & m
\end{array}
\right] .  \label{wavelet}
\end{equation}
The symplectic transformation will be

\begin{equation}
S=\frac{1}{\sqrt{2}}\left[
\begin{array}{ll}
s_{1} & s_{2} \\
s_{1} & -s_{2}
\end{array}
\right] \oplus \frac{1}{\sqrt{2}}\left[
\begin{array}{ll}
1/s_{1} & 1/s_{2} \\
1/s_{1} & -1/s_{2}
\end{array}
\right] ,
\end{equation}
where $s_{1}=\left( \frac{m+k_{q}}{m-k_{p}}\right) ^{\frac{1}{4}},$ $%
s_{2}=\left( \frac{m-k_{q}}{m+k_{p}}\right) ^{\frac{1}{4}}.$ And one obtains
the M-matrix for the state

\begin{equation}
M=(S^T)^{-1}\log \frac{2\widetilde{\alpha }+1}{2\widetilde{\alpha }-1}S^{-1}
\label{wavelet1}
\end{equation}
where $\widetilde{\alpha }=\widetilde{\alpha }_q\oplus \widetilde{\alpha }_p$%
, $\widetilde{\alpha }_q=\widetilde{\alpha }_p=\frac 12diag\{\sqrt{\left(
m+k_q\right) \left( m-k_p\right) },\sqrt{\left( m-k_q\right) \left(
m+k_p\right) }\}$.

The above transform is limited to the $q-p$ decorrelation matrices. For a
more general CM to EM transform, although the symplectic transformation is
also available up to the length and phase factor of each eigenvector of \ $%
\Delta ^{-1}\alpha $, the symplectic condition $S\Delta S^{T}=\Delta $ is
not easily verified. So let us turn to a direct way of transforming CM to EM
as well as EM to CM. This is due to the fact that $\alpha =S\widetilde{%
\alpha }S^{T}$ and $S^{T}MS=$ $\widetilde{M}$. The late can be rewritten as $%
M^{-1}=S$ $\widetilde{M}^{-1}S^{T}.$ So that $\alpha $ and $M^{-1}$can be
simultaneously symplectic diagonalized. $\Delta ^{-1}\alpha $ and $\Delta
^{-1}M^{-1}$ will have the common eigenfunctions. And they commutate with
each other, $\left[ \Delta ^{-1}\alpha ,\Delta ^{-1}M^{-1}\right] =0.$ Hence
\begin{equation}
M\alpha \Delta ^{-1}-\Delta ^{-1}\alpha M=0.
\end{equation}
Together with $\widetilde{\alpha }=\frac{1}{2}\coth \frac{1}{2}\widetilde{M}%
, $ one can transform $\alpha $ to $M$ or $M$ to $\alpha $.

\section{GREE and border state}

Now given a CM, one can transform it into EM. This enables the calculation
of relative entropy between two gaussian states to be practically possible.
The relative entropy of a gaussian state $\rho $ with respect to another
gaussian state $\sigma $ is defined as
\begin{equation}
S\left( \rho \left\| \sigma \right. \right) =Tr\rho (\log \rho -\log \sigma
).
\end{equation}
The normalization factor of the state $\sigma $ is

\begin{equation}
c=\prod_{j=1}^{n}2\sinh \frac{\widetilde{M}_{\sigma j}}{2}=\prod_{j=1}^{n}%
\frac{1}{\sqrt{\gamma _{\sigma j}^{2}-\frac{1}{4}}}.
\end{equation}
Hence

\begin{eqnarray}
-Tr\rho \log \sigma &=&-\log c+\frac{1}{2}Tr\rho F^{T}M_{\sigma }F  \notag \\
&=&-\log c+\frac{1}{2}Tr\left( \alpha _{\rho }-\frac{i}{2}\Delta \right)
M_{\sigma }  \notag \\
&=&-\log c+\frac{1}{2}Tr\alpha _{\rho }M_{\sigma },
\end{eqnarray}
where the operator trace $TrF\rho F^{T}=\left( \alpha _{\rho }-\frac{i}{2}%
\Delta \right) $ \cite{Holevo1} and the fact that $Tr\Delta M_{\sigma }=$ $%
Tr\Delta (S^{T})^{-1}\widetilde{M}_{\sigma }S^{-1}$ $\ =TrS^{-1}\Delta
(S^{T})^{-1}\widetilde{M}_{\sigma }=Tr\Delta \widetilde{M}_{\sigma }=0$ have
been used. The trace in the first equality is two fold, operator trace and
matrix trace.

The relative entropy of entanglement was defined as the minimization of the
relative entropy of a state with respect to all separable state: $E_R(\rho
)=\min_{\sigma \in D}S(\rho \left\| \sigma \right. ),$ where $D$ is the set
of separable state. If its subset $D_G$ of all gaussian state is used
instead of the set $D$ itself , then the GREE for a state can be defined as :

\begin{equation}
E_{GR}(\rho )=\min_{\sigma \in D_{G}}S(\rho \left\| \sigma \right. ).
\end{equation}
I will prove that the separable set can be further restricted to the border
separable set. For completeness I will start theorem 1.

Theorem 1: \textit{The relative entropy of entanglement is obtained when the
separable state is at the border of the set of separable states and the set
of inseparable states.}

Proof: The relative entropy is jointly convex in its arguments \cite{Leib}.
That is, if $\rho _{1}$, $\rho _{2}$, $\sigma _{1}$ and $\sigma _{2}$ are
density operators, and $p_{1}$ and $p_{2}$ are non-negative numbers that sum
to unity (i.e., probabilities), then $S\left( \rho \left\| \sigma \right.
\right) \leq p_{1}S\left( \rho _{1}\left\| \sigma _{1}\right. \right)
+p_{2}S\left( \rho _{2}\left\| \sigma _{2}\right. \right) ,$where $\rho
=p_{1}\rho _{1}+p_{2}\rho _{2}$, and $\sigma $ = $p_{1}\sigma
_{1}+p_{2}\sigma _{2}$. Joint convexity automatically implies convexity in
each argument, so that

\begin{equation}
S\left( \rho \left\| \sigma \right. \right) \leq p_{1}S\left( \rho \left\|
\sigma _{1}\right. \right) +p_{2}S\left( \rho \left\| \sigma _{2}\right.
\right) .
\end{equation}
and for $0\leq x\leq 1,$ one has $S\left( \rho \left\| \left( 1-x\right)
\rho +x\sigma \right. \right) \leq (1-x)S\left( \rho \left\| \rho \right.
\right) +xS\left( \rho \left\| \sigma \right. \right) =xS\left( \rho \left\|
\sigma \right. \right) \leq S\left( \rho \left\| \sigma \right. \right) .$
Hence for a separable state $\sigma $ that is not at the border one can find
a new separable state with less relative entropy until the new separable
state is at the border.

Theorem 2: \textit{The gaussian relative entropy of entanglement for
gaussian state is obtained when the gaussian separable state is at the
border of the set of separable states and the set of inseparable states.}

Proof: The idea is like this: for any given separable gaussian state $\sigma
_{0}$, one needs to find a line to connect $\sigma _{0}$ and the inseparable
gaussian state $\rho $ with every point in the line is a gaussian state
which is denoted as $\sigma $. In the line, between the separable state $%
\sigma _{0}$ and inseparable state $\rho $, there should be a border
gaussian state. I will find such a line by continuously change the state $%
\sigma $ in the fashion that the relative entropy of $\rho $ with respect to
$\sigma $ decreases monotonically. If the process of decreasing of relative
entropy does not stop, then the relative entropy will go to its minimum
value. Because $S\left( \rho \left\| \sigma \right. \right) \geq 0$ and with
equality iff $\rho =\sigma $, so $\sigma $ will eventually reach $\rho $. In
the following I will mainly prove that the decreasing process would not stop
if $\sigma \neq \rho $.

Now unitary operations leave $S\left( \rho \left\| \sigma \right. \right) $
invariant, i.e. $S\left( \rho \left\| \sigma \right. \right) =S\left( U\rho
U^{+}\left\| U\sigma U^{+}\right. \right) .$ This reflects the fact that
\begin{equation}
Tr\alpha _\rho M_\sigma =Tr\alpha _\rho \left( S_\sigma ^T\right) ^{-1}%
\widetilde{M}_\sigma S_\sigma ^{-1}=TrS_\sigma ^{-1}\alpha _\rho \left(
S_\sigma ^T\right) ^{-1}\widetilde{M}_\sigma .
\end{equation}
Denote $\beta =S_\sigma ^{-1}\alpha _\rho \left( S_\sigma ^T\right) ^{-1}$.
The relative entropy will be

\begin{equation}
S\left( \rho \left\| \sigma \right. \right) =Tr\rho \log \rho
-\sum_{j=1}^{n}\log \left( 2\sinh \frac{\widetilde{M}_{\sigma j}}{2}\right) +%
\frac{1}{2}\sum_{j=1}^{n}\left( \beta _{jj}+\beta _{n+j,n+j}\right)
\widetilde{M}_{\sigma j}.
\end{equation}
Because $\beta $ is also a CM of some gaussian state, the uncertainty
relation requires $\beta -\frac{i}{2}\Delta \geq 0$. Hence $\beta _{jj}\beta
_{n+j,n+j}\geq \frac{1}{4}$, and $\frac{1}{2}\left( \beta _{jj}+\beta
_{n+j,n+j}\right) \geq \sqrt{\beta _{jj}\beta _{n+j,n+j}}\geq \frac{1}{2}.$
Denote $\overline{\beta }_{jj}=\frac{1}{2}\left( \beta _{jj}+\beta
_{n+j,n+j}\right) $. The partial derivatives of the relative entropy are

\begin{eqnarray}
\frac{\partial S\left( \rho \left\| \sigma \right. \right) }{\partial
\widetilde{M}_{\sigma j}} &=&\frac{1}{2}\left( \beta _{jj}+\beta
_{n+j,n+j}\right) -\frac{1}{2}\coth \frac{\widetilde{M}_{\sigma j}}{2}
\notag \\
&=&\overline{\beta }_{jj}-\gamma _{\sigma j}.
\end{eqnarray}
Because $\widetilde{M}_{\sigma j}$ is a monotonically decreasing function of
$\gamma _{\sigma j}$, the partial derivative of the relative entropy with
respect to $\gamma _{\sigma j}$ is positive iff $\gamma _{\sigma j}>%
\overline{\beta }_{jj}$.\ The line designed to connect $\sigma _{0}$ and $%
\rho $ is like so: first let us fix $S_{\sigma }$ to $S_{\sigma _{0}}$, then
fix all $\gamma _{\sigma j}$ to $\gamma _{\sigma _{0}j}$ except $\gamma
_{\sigma 1}$, if $\gamma _{\sigma 1}>\overline{\beta }_{11}$, then decrease $%
\gamma _{\sigma 1}$ until it is equal to $\overline{\beta }_{11}$, the
relative entropy decreases monotonically. if $\gamma _{\sigma 1}<\overline{%
\beta }_{11}$, then increase $\gamma _{\sigma 1}$ until it is equal to $%
\overline{\beta }_{11}$, the relative entropy also decreases monotonically.
Now the state is with $\gamma _{\sigma 1}=$ $\overline{\beta }_{11}$ and all
other $\gamma _{\sigma j}=$ $\gamma _{\sigma _{0}j}$. Then let us make $%
\gamma _{\sigma 2}$ monotonically vary to $\overline{\beta }_{22}$ while
keeping the other $\gamma _{\sigma j}$ fixed. At last all $\gamma _{\sigma
j}=\overline{\beta }_{jj}$, and in every step the relative entropy decreases
monotonically. Now the state $\sigma $ is with its all symplectic
eigenvalues $\gamma _{\sigma j}=\overline{\beta }_{jj}$ but with $S_{\sigma
} $ still being fixed to $S_{\sigma _{0}}$. The relative entropy then will
be
\begin{equation}
S\left( \rho \left\| \sigma \right. \right) =Tr\rho \log \rho
+\sum_{j=1}^{n}g\left( \overline{\beta }_{jj}-\frac{1}{2}\right) ,
\label{www}
\end{equation}
where $g\left( x\right) =\left( x+1\right) \log \left( x+1\right) -x\log x$
is the bosonic entropy function, which is a monotonically increase function
of its argument, but its derivative $\frac{dg\left( x\right) }{dx}$
decreases with $x$ increases.

The next step of stretching the line is to change $S_{\sigma }$ gradually in
order that $S\left( \rho \left\| \sigma \right. \right) $decreases further.
Now $\beta =S_{\sigma }^{-1}S_{\rho }\widetilde{\alpha }_{\rho }S_{\rho
}^{T}\left( S_{\sigma }^{T}\right) ^{-1}$, let us apply infinitive small
symplectic transform to $\beta $ to change the state $\sigma $ continuously.
The infinitive small symplectic transforms will accumulate some finite
symplectic transforms, which are the following six kinds: (i) Local
rotations which keep $\overline{\beta }_{jj}$ invariant; (ii)Local
squeezings which can be used to decrease $\overline{\beta }_{jj}$ to $\sqrt{%
\beta _{jj}\beta _{j+n,j+n}}$ ; (iii) The first kind two mode rotations. For
modes $i$ and $j$, if the two mode CM $\beta _{sub}$ (submatrix of $\beta $)
is arranged according to the order of canonical operators $\left[
q_{i},q_{j};p_{i},p_{j}\right] $, the rotation will be $\Theta \left( \theta
\right) \oplus \Theta \left( \theta \right) $, where
\begin{equation}
\Theta \left( \theta \right) =\left[
\begin{array}{ll}
\cos \theta & \sin \theta \\
-\sin \theta & \cos \theta
\end{array}
\right] .
\end{equation}
Before the rotation is applied, $\beta _{sub}$ has already been prepared
(local squeezed) to the form of equal diagonal elements within each mode,
that is $\overline{\beta }_{ii}=\beta _{ii}=\beta _{i+n,i+n}$ and the same
for mode $j$. The inter-mode rotation keeps $\overline{\beta }_{ii}+%
\overline{\beta }_{jj}$ that is the trace of the two mode CM
$\beta _{sub}$ invariant. The only way to decrease the relative
entropy is to enlarge the difference between $\overline{\beta
}_{ii}$ and $\overline{\beta }_{jj}$. This is because that the
bigger one say $\overline{\beta }_{ii}$ increases some amount, the
smaller one $\overline{\beta }_{jj}$ will decrease the same
amount, but total relative entropy will decrease due to the
monotonically
decreasing property of the derivative of bosonic entropy function $\frac{%
dg\left( x\right) }{dx}$. The distance between $\overline{\beta }_{ii}$ and $%
\overline{\beta }_{jj}$ can be enlarged at most to $\left| \overline{\beta }%
_{ii}^{^{\prime }}-\overline{\beta }_{jj}^{^{\prime }}\right| =\sqrt{\left|
\overline{\beta }_{ii}-\overline{\beta }_{jj}\right| ^{2}+\left( \beta
_{ij}+\beta _{i+n,.j+n}\right) ^{2}}$ by proper rotation. After such
rotation,we have $\beta _{ij}^{^{\prime }}=-\beta _{i+n,.j+n}^{^{\prime }}$,
the off diagonal elements of $q$ part and $p$ part are asymmetrized ;(iv)
The first kind two mode squeezings $R\left( r\right) \oplus R\left(
-r\right) $, where
\begin{equation}
R\left( r\right) =\left[
\begin{array}{ll}
\cosh r & \sinh r \\
\sinh r & \cosh r
\end{array}
\right] .
\end{equation}
After successive applying of (ii) and (iii) for rounds, (numeric results
indicate the a few rounds will do), $\beta _{sub}$ will have the form with $%
\beta _{ii}=\beta _{i+n,i+n}$, $\beta _{jj}=\beta _{j+n,j+n}$,$\beta
_{ij}=-\beta _{i+n,.j+n}$, The two mode squeezing then will be used to
diagonalize this part of $\beta _{sub}$ matrix. The squeezing decreases $%
\overline{\beta }_{ii}$ and $\overline{\beta }_{jj}$ with the same amount
which can be at most $\frac{1}{2}(\overline{\beta }_{ii}+\overline{\beta }%
_{jj})-\sqrt{\frac{1}{4}(\overline{\beta }_{ii}+\overline{\beta }%
_{jj})^{2}-\beta _{ij}^{2}}$. (v) The second kind two mode rotations which
rotate $q_{i},p_{j}$ pair and simultaneously $p_{i},q_{j}$ pair. The
rotation is similar to the first kind two mode rotation but with distance
between $\overline{\beta }_{ii}$ and $\overline{\beta }_{jj}$ can be
enlarged at most to $\left| \overline{\beta }_{ii}^{^{\prime }}-\overline{%
\beta }_{jj}^{^{\prime }}\right| =\sqrt{\left| \overline{\beta }_{ii}-%
\overline{\beta }_{jj}\right| ^{2}+\left( \beta _{i,j+n}-\beta
_{i+n,.j}\right) ^{2}}$. And one has $\beta _{i,j+n}^{^{\prime }}=\beta
_{i+n,.j}^{^{\prime }}$ after the such rotation. (vi) The second kind two
mode squeezings which squeeze $q_{i},p_{j}$ pair and simultaneously $%
p_{i},q_{j}$ pair. After successive applying of (ii) and (v) for rounds, $%
\beta _{sub}$ will have the form with $\beta _{ii}=\beta _{i+n,i+n}$, $\beta
_{jj}=\beta _{j+n,j+n}$, $\beta _{i,j+n}=\beta _{i+n,.j}$, The two mode
squeezing then will be used to diagonalize this part of $\beta _{sub}$
matrix. The squeezing decreases $\overline{\beta }_{ii}$ and $\overline{%
\beta }_{jj}$ with the same amount which can be at most $\frac{1}{2}(%
\overline{\beta }_{ii}+\overline{\beta }_{jj})-\sqrt{\frac{1}{4}(\overline{%
\beta }_{ii}+\overline{\beta }_{jj})^{2}-\beta _{i,j+n}^{2}}$.
These six symplectic transforms are classified as three group:
(i); (ii)-(iii)-(iv); (ii)-(v)-(vi). Each group aims at
diagonalized 4 of the off-diagonal elements of the two mode CM
while decreasing the relative entropy except the group (i) which
keeps the relative entropy. By successively apply the three groups
the relative entropy will decrease step by step before it is
diagonalized. Then the whole procedure of diagonalizing is applied
to all other pairs of modes round and round. Before $\beta $ is
totally diagonalized, a way can always\ be found to decrease the
relative entropy. The totally diagonalized $\beta $ (denoted as
$\widetilde{\beta }$) is exactly $\widetilde{\alpha }$. This is
because the original $\beta =S_{\sigma }^{-1}\alpha _{\rho }\left(
S_{\sigma }^{T}\right) ^{-1}$ has the same symplectic eigenvalues
with $\alpha _{\rho }$, so $\widetilde{\beta }$ and
$\widetilde{\alpha }$ may only differ by the interchange of mode,
However the relative entropy is $0$ according to Eq.(\ref{www}).
This is only possible when $\sigma =\rho $. So that $\sigma $ at
last reaches $\rho . $ Maybe the gradually changing $\sigma $
passes the border of separable and inseparable sets several times,
but this does not matter, the last passing meets the requirement
of GREE.\ And theorem 2 is proved.

In finding the minimization in the border state set, one has another
question, that is, if displacement decrease the relative entropy or not? The
answer is negative. The operator is $\exp \left( iF^{T}z\right) ,$ where $z$
is a real $2n$ vector. Since

\begin{equation}
\exp \left( iF^Tz\right) \sigma \exp \left( -iz^TF\right) =c\exp [-\frac
12\left( F+\Delta z\right) ^TM_\sigma \left( F+\Delta z\right) ]
\end{equation}

\begin{eqnarray}
-Tr\rho \log \sigma &=&-\log c+\frac{1}{2}Tr\rho \left( F+\Delta z\right)
^{T}M_{\sigma }\left( F+\Delta z\right)  \notag \\
&=&-\log c+\frac{1}{2}Tr\alpha _{\rho }M_{\sigma }+\frac{1}{2}Tr\left[
\left( \Delta z\right) ^{T}M_{\sigma }\Delta z\right] ,
\end{eqnarray}
While $M_{\sigma }$ is positive definite (I will elucidate it at section 5),
so the last term is not less than $0$. The displacement can not decrease the
relative entropy.

To find the GREE of state $\rho $ is to find the $M_\sigma $ matrix of a
border state $\sigma $ such that the relative entropy reaches its minimum.

\section{GREE of $1\times 1$ gaussian state system}

Now let us turn to the $1\times 1$ gaussian state system. The general case
of relative entropy is that $\alpha _{\rho }$ is in its standard form but $%
M_{\sigma }$ is not. It is no need to require that they are all in the most
general form, because by the unitary invariant of relative entropy, at least
one of the matrices $\alpha _{\rho }$ and $M_{\sigma }$ can be converted to
any possible form. Since any possible $M_{\sigma }$ matrix can be simplified
to its standard form by local operations\cite{Duan}, so such a $M_{\sigma }$%
can be generated from the standard form $M_{\sigma s}$. For $1\times 1$
system, suppose $\alpha _{\rho }$ takes the standard form,

\begin{equation}
\alpha _{\rho q}=\left[
\begin{array}{cc}
\alpha _{1} & \alpha _{2} \\
\alpha _{2} & \alpha _{3}
\end{array}
\right] ,\text{ \ \ }\alpha _{\rho q}=\left[
\begin{array}{cc}
\alpha _{1} & \alpha _{4} \\
\alpha _{4} & \alpha _{3}
\end{array}
\right] ,
\end{equation}
$M_{\sigma s}$ takes the same form but with elements $M_{si}$ $(i=1,\cdots
,4)$ respectively. The local operations are firstly a local rotation $L_{1}$
with angles $\theta _{A1}$ and $\theta _{B1}$ for the two modes
respectively, then a local squeezing $L_{2}=diag\{\exp \left( \tau
_{A}\right) ,\exp \left( \tau _{B}\right) ,\exp \left( -\tau _{A}\right)
,\exp \left( -\tau _{B}\right) \}$, then another local rotation $L_{3}$ with
angles $\theta _{A2}$ and $\theta _{B2}$ for the two modes respectively. The
standard form of $M_{\sigma s}$ is modified to $M_{\sigma
}=L_{3}^{T}L_{2}^{T}L_{1}^{T}M_{\sigma s}L_{1}L_{2}L_{3}$. Local operations
will leave the normalization factor $c$ unchanged, so one just needs to
consider $Tr\alpha _{\rho }M_{\sigma }$ term, one gets

\begin{eqnarray}
Tr\alpha _\rho M_\sigma &=&2\alpha _1M_{s1}\cosh 2\tau _A+2\alpha
_3M_{s3}\cosh 2\tau _B  \notag \\
&&+\cos \left( \theta _{A1}+\theta _{B1}-\theta _{A2}-\theta _{B2}\right)
\left( \alpha _2-\alpha _4\right) \left( M_{s2}-M_{s4}\right) \sinh \tau
_A\sinh \tau _B  \notag \\
&&+\cos \left( \theta _{A1}+\theta _{B1}+\theta _{A2}+\theta _{B2}\right)
\left( \alpha _2-\alpha _4\right) \left( M_{s2}-M_{s4}\right) \cosh \tau
_A\cosh \tau _B  \notag \\
&&+\cos \left( \theta _{A1}+\theta _{B1}-\theta _{A2}+\theta _{B2}\right)
\left( \alpha _2-\alpha _4\right) \left( M_{s2}-M_{s4}\right) \sinh \tau
_A\sinh \tau _B  \notag \\
&&+\cos \left( \theta _{A1}+\theta _{B1}+\theta _{A2}-\theta _{B2}\right)
\left( \alpha _2+\alpha _4\right) \left( M_{s2}-M_{s4}\right) \cosh \tau
_A\sinh \tau _B  \notag \\
&&+\cos \left( \theta _{A1}-\theta _{B1}-\theta _{A2}-\theta _{B2}\right)
\left( \alpha _2-\alpha _4\right) \left( M_{s2}+M_{s4}\right) \sinh \tau
_A\cosh \tau _B  \notag \\
&&+\cos \left( \theta _{A1}-\theta _{B1}+\theta _{A2}+\theta _{B2}\right)
\left( \alpha _2-\alpha _4\right) \left( M_{s2}+M_{s4}\right) \cosh \tau
_A\sinh \tau _B  \notag \\
&&+\cos \left( \theta _{A1}-\theta _{B1}-\theta _{A2}+\theta _{B2}\right)
\left( \alpha _2+\alpha _4\right) \left( M_{s2}+M_{s4}\right) \sinh \tau
_A\sinh \tau _B  \notag \\
&&+\cos \left( \theta _{A1}-\theta _{B1}+\theta _{A2}-\theta _{B2}\right)
\left( \alpha _2+\alpha _4\right) \left( M_{s2}+M_{s4}\right) \cosh \tau
_A\cosh \tau _B.
\end{eqnarray}
Clearly, in order that $Tr\alpha _\rho M_\sigma $ is minimized, the local
rotations should be arranged in such a way that all the $\cos $-factors are $%
\pm 1,$ then

\begin{eqnarray}
Tr\alpha _{\rho }M_{\sigma } &=&2\alpha _{1}M_{s1}\cosh 2\tau _{A}+2\alpha
_{3}M_{s3}\cosh 2\tau _{B}  \notag \\
&&-\left[ \left| \left( \alpha _{2}-\alpha _{4}\right) \left(
M_{s2}-M_{s4}\right) \right| +\left| \left( \alpha _{2}+\alpha _{4}\right)
\left( M_{s2}+M_{s4}\right) \right| \right] \cosh \left( \left| \tau
_{A}\right| +\left| \tau _{B}\right| \right)  \notag \\
&&-\left[ \left| \left( \alpha _{2}-\alpha _{4}\right) \left(
M_{s2}+M_{s4}\right) \right| +\left| \left( \alpha _{2}+\alpha _{4}\right)
\left( M_{s2}-M_{s4}\right) \right| \right] \sinh \left( \left| \tau
_{A}\right| +\left| \tau _{B}\right| \right) .
\end{eqnarray}
Without lose of generality, let $\alpha _{2}>0,\alpha _{4}<0$ and $\alpha
_{2}>-\alpha _{4},$ and the $M_{\sigma }$ matrix now has the form of $%
L_{2}^{\prime }M_{\sigma s}^{\prime }L_{2}^{\prime }$ with $L_{2}^{\prime
}=diag\{\exp \left| \tau _{A}\right| ,\exp \left| \tau _{B}\right| ,\exp
\left( -\left| \tau _{A}\right| \right) ,\exp \left( -\left| \tau
_{B}\right| \right) \}$. $M_{\sigma s}^{\prime }$ only differs from $%
M_{\sigma s}$ by off diagonal elements. For simplification of the notations,
denote $M_{\sigma si}^{\prime }$ as $M_{i}$, then $M_{1(3)}=M_{s1(3)}$ and $%
M_{2(4)}=$ $-\frac{1}{2}(\left| M_{s2}+M_{s4}\right| $ $\pm \left|
M_{s2}-M_{s4}\right| ).$

The problem now is to determine the elements of the matrix $M_{\sigma }$ of
a border state $\sigma $. The local squeezing $L_{2}^{\prime }$ can be
rewritten as the product of two local squeezings $Y$ and $X$, Now $M_{\sigma
}=Y(y)X(x)M_{\sigma s}^{\prime }X(x)Y(y)$, with $Y(y)=diag\{\sqrt{y},\sqrt{y}%
,\sqrt{y}^{-1},\sqrt{y}^{-1}\}$ and $X(x)=diag\{\sqrt{x},\sqrt{x}^{-1},\sqrt{%
x}^{-1},\sqrt{x}\}$ are symplectical transformations. After minimization of $%
Tr\alpha _{\rho }M_{\sigma }$ with respect to $y$, one has

\begin{equation}
\frac{1}{2}Tr\alpha _{\rho }M_{\sigma }=\sqrt{(\alpha _{1}M_{1}x+\alpha
_{3}M_{3}x^{-1}+2\alpha _{2}M_{2})(\alpha _{1}M_{1}x^{-1}+\alpha
_{3}M_{3}x+2\alpha _{4}M_{4})}.
\end{equation}
The further minimization of $\frac{1}{2}Tr\alpha _{\rho }M_{\sigma }$ with
respect to $x$ will lead to an algebra equation of $x$ up to power of $4$.
Although after this round of minimization, $\frac{1}{2}Tr\alpha _{\rho
}M_{\sigma }$ can not easily be expressed, but in principle it is possible
to analytically express it as a function of the four $M_{i}$. The GREE
problem of gaussian state $\rho $ now is the minimization of $-\log c+\frac{1%
}{2}Tr\alpha _{\rho }M_{\sigma }$ where the whole function now has only the
four $M_{i}$ as its variables. The four $M_{i}$ would fulfill the border
state condition, so there are only $3$ of them left to be determined in
further minimization of the relative entropy. Consider the border state
condition for state characterized by its standard form $M_{\sigma s}^{\prime
}$, that is by the four $M_{i}$, suppose the corresponding CM is $\alpha
=\alpha _{q}\oplus \alpha _{p}$, the condition for the state to be a border
state is \cite{Simon}:
\begin{equation}
4\det \left( \alpha _{q}\alpha _{p}\right) =Tr\left( \alpha _{q}\alpha
_{p}\right) +2\left( \left| c_{1}c_{2}\right| -c_{1}c_{2}\right) -\frac{1}{4}%
.  \label{ww0}
\end{equation}
where $c_{1}$, $c_{2}$ are off diagonal elements of $\alpha _{q}$ and $%
\alpha _{p}$ respectively. By Eqs. (\ref{wave0}), one has $\det \left(
\alpha _{q}\alpha _{p}\right) =\gamma _{A}^{2}\gamma _{B}^{2}$ and
\begin{equation}
Tr\left( \alpha _{q}\alpha _{p}\right) =\gamma _{A}^{2}+\gamma _{B}^{2},
\label{ww1}
\end{equation}
where $\gamma _{A},\gamma _{B}$ are the symplectic eigenvalue of $\alpha $
for the two modes respectively, and $\gamma _{j}=\frac{1}{2}\coth \frac{%
\widetilde{M}_{j}}{2},$ $(j=A,B)$. While $\widetilde{M}_{j}$ are the
symplectic eigenvalues of $M_{\sigma s}^{\prime }$, what left is to
determine $c_{1}c_{2}$. The commutation relation $\left[ \Delta ^{-1}\alpha
,\Delta ^{-1}M^{-1}\right] =0$ now takes the form $M_{p}\alpha _{p}=\alpha
_{q}M_{q}$ or equivalently $\alpha _{p}M_{p}=M_{q}\alpha _{q}$. This enables
all the other elements of $\alpha $ to be expressed as linear combination of
$c_{1}$ and $c_{2}$. Since $\det \left( \alpha _{q}\alpha _{p}\right) =\det
\left( \alpha _{q}M_{p}^{-1}\alpha _{q}M_{q}\right) =$ $\left[ \det \left(
\alpha _{q}\right) \right] ^{2}\det \left( M_{q}\right) /\det \left(
M_{p}\right) ,$ so that
\begin{equation}
\det \left( \alpha _{q}\right) =\gamma _{A}\gamma _{B}\sqrt{\det \left(
M_{p}\right) /\det \left( M_{q}\right) }.  \label{ww2}
\end{equation}
Either $Tr\left( \alpha _{q}\alpha _{p}\right) $ or $\det \left( \alpha
_{q}\right) $ is a linear combination of $c_{1}^{2}$, $c_{2}^{2}$ and $%
c_{1}c_{2}.$ By combining Eq.(\ref{ww1}) and Eq.(\ref{ww2}), one arrives at
a quadratic equation about $c_{1}c_{2}.$ So that $c_{1}c_{2}$ can be
expressed with $M_{j}$ so does the border state condition of Eq.(\ref{ww0}).

In this section, I reduced the EM of the destination state from $10$
parameters to $3$ ($4$ parameters and $1$ restriction, strictly speaking).
In the next section I will elucidate that it is really $3.$ These $3$
parameters are left for numeric calculation, because minimization function
at this step is too complicated to be dealt with analytically.

\section{Classification of $1\times 1$ gaussian states}

In this section, I will aim at constructing the six parameter $M_{\sigma }$,
the footnote $\sigma $ will\ be omitted when it is not confusing. The CM $%
\alpha $ should satisfy uncertainty relation $\alpha -\frac{i}{2}\Delta \geq
0$. The standard form of the correlation matrix of 1$\times $1 gaussian
state have four parameters. Uncertainty relation adds some restrictions
among the four parameters. So that the parameters are not freely chosen,
otherwise the state may not be physical. Needlessly to say, EM $M$ is less
restricted than CM $\alpha $. If all symplectic eigenvalues of $M$ is
positive, the state should be physical, because the density operator will be
in the form of $\sigma \sim U\exp (-\frac{1}{2}\sum_{j}\widetilde{M}%
_{j}(q_{j}^{2}+p_{j}^{2}))U^{+}=$ $AA^{+},$ with $A=U\exp (-\frac{1}{4}%
\sum_{j}\widetilde{M}_{j}(q_{j}^{2}+p_{j}^{2}))$, so that $\sigma $ is
positive definite. For standard form $M$ of $1\times 1$ gaussian state, it
is easy to check that $M$ should be positive definite. But as elucidated in
the former section, the separable criterion is quite complicate expressed
with $M$. In this section, I will seek free parameter representations which
are simple both in the uncertainty relation and separable criterion.

Given a standard form of $\alpha =\alpha _{q}\oplus \alpha _{p},$ with
\begin{equation}
\alpha _{q}=\left[
\begin{array}{ll}
a & c_{1} \\
c_{1} & b
\end{array}
\right] ,\text{ \qquad }\alpha _{p}=\left[
\begin{array}{ll}
a & -c_{2} \\
-c_{2} & b
\end{array}
\right] ,
\end{equation}
where without lose of generality $c_{1}>0$ and $-c_{2}<0$ are supposed (
when $c_{1}(-c_{2})\geq 0$, the state is definitely separable, so that will
not be the border state of interesting). Now let us seek an operational way
to symplectically diagonalize it. This is accomplished by first applying
local squeezing $X$, then a two mode squeezing or a two mode rotation
according to different structure of $\alpha .$ The CM will be transformed to
(i) $R\left( r\right) \oplus R\left( -r\right) X(x)\alpha X(x)R\left(
r\right) \oplus R\left( -r\right) $ or (ii) $\Theta \left( \theta \right)
\oplus \Theta \left( \theta \right) X(x)\alpha X(x)\Theta \left( -\theta
\right) \oplus \Theta \left( -\theta \right) .$ In case (i), $x$ is so
chosen such that $(ax+bx^{-1})/c_{1}=(ax^{-1}+bx)/c_{2}$, $x=\sqrt{\frac{%
ac_{1}-bc_{2}}{ac_{2}-bc_{1}}}$, then a two mode squeezing with $\tanh (2r)$
$=-2\sqrt{(ac_{1}-bc_{2})(ac_{2}-bc_{1})}/\left| a^{2}-b^{2}\right| $ will
diagonalize the properly local squeezed CM. Clearly the existence of $x$
requires that $\frac{a}{b}+\frac{b}{a}>\frac{c_{1}}{c_{2}}+\frac{c_{2}}{c_{1}%
}$. In case (ii), $x$ is so chosen such that $(ax-bx^{-1})/c_{1}$ $%
=-(ax^{-1}-bx)/c_{2}$, $x=\sqrt{\frac{ac_{1}-bc_{2}}{bc_{1}-ac_{2}}}$, then
a two mode rotation with $\tan (2\theta )$ $=2\sqrt{%
(ac_{1}-bc_{2})(bc_{1}-ac_{2})}/(a^{2}-b^{2})$ $\ \cdot $ $%
sign(bc_{1}-ac_{2})$ will diagonalized the CM. The existence of $x$ requires
that $\frac{a}{b}+\frac{b}{a}<\frac{c_{1}}{c_{2}}+\frac{c_{2}}{c_{1}}$.
After the diagonalization of these two cases, further local squeezing of $%
X(x^{\prime })$ and $Y(y)$ will be applied to transform their diagonal
elements into symplectic eigenvalues. Certainly there are the third case of $%
\frac{a}{b}+\frac{b}{a}=\frac{c_{1}}{c_{2}}+\frac{c_{2}}{c_{1}}$ and the
fourth case of $a=b$ which is the case of symmetric gaussian states. So,
according to $\frac{a}{b}+\frac{b}{a}\ $being more than or less than or
equal to $\frac{c_{1}}{c_{2}}+\frac{c_{2}}{c_{1}}$, the state is classified
as type (i), type (ii), type (iii). And if $a=b$, the state will be type
(iv) state. The quantity\ $(\frac{a}{b}+\frac{b}{a})/(\frac{c_{1}}{c_{2}}+%
\frac{c_{2}}{c_{1}})$ is critical in symplectic diagonalization of the CM.
It is a kind of ratio of diagonal elements to off diagonal elements of the
CM.

Now let us construct the CM of all four classes. The process is just the
reverse of the diagonalization. Let us begin with $\widetilde{\alpha }%
=diag\{\gamma _{A},\gamma _{B};\gamma _{A},\gamma _{B}\}$, Then $X(x^{\prime
-1})$ is applied. $Y(y^{-1})$ can be put at the last stage because it
commutates with all other kind operations of $q-p$ decorrelation type such
as the two mode rotation and the two mode squeezing and local squeezing $X$.
The next step is to apply $R\left( -r\right) \oplus R\left( r\right) $ or $%
\Theta \left( -\theta \right) \oplus \Theta \left( -\theta \right) $ for
type (i) and type (ii) CMs or states respectively. The separable property is
totally determined after this step and the successively applications of
local squeezing operations $X(x^{-1})$ and $Y(y^{-1})$ will not affect the
separability. So when the separability is concerned, the last two local
squeezing operations can be omitted. The CM generated will be (i):$R\left(
-r\right) \oplus R\left( r\right) X(x^{\prime -1})\widetilde{\alpha }%
X(x^{\prime -1})R\left( -r\right) \oplus R\left( r\right) $ and (ii) $\Theta
\left( -\theta \right) \oplus \Theta \left( -\theta \right) X(x^{\prime -1})%
\widetilde{\alpha }X(x^{\prime -1})\Theta \left( \theta \right) \oplus
\Theta \left( \theta \right) .$ The separable criterion for the two cases
then will be
\begin{equation}
(2\gamma _{A}^{2}-\frac{1}{2})(2\gamma _{B}^{2}-\frac{1}{2})\geq \sinh
^{2}(2r)[\left( x^{\prime 2}+x^{\prime -2}\right) \gamma _{A}\gamma
_{B}+(\gamma _{A}^{2}+\gamma _{B}^{2})]  \label{ww3}
\end{equation}
for type (i) and
\begin{equation}
(2\gamma _{A}^{2}-\frac{1}{2})(2\gamma _{B}^{2}-\frac{1}{2})\geq \sin
^{2}(2\theta )[\left( x^{\prime 2}+x^{\prime -2}\right) \gamma _{A}\gamma
_{B}-(\gamma _{A}^{2}+\gamma _{B}^{2})]  \label{ww4}
\end{equation}
with $x^{\prime }>\max \{\sqrt{\frac{\gamma _{A}}{\gamma _{B}}},\sqrt{\frac{%
\gamma _{B}}{\gamma _{A}}}\}$ or $x^{\prime }<\min \{\sqrt{\frac{\gamma _{A}%
}{\gamma _{B}}},\sqrt{\frac{\gamma _{B}}{\gamma _{A}}}\}$ for type (ii),
where the equalities in these two equation are for the border states, For
border state, one of the parameters, say $x^{\prime },$ can be easily
expressed by the other three parameters. The corresponding EMs will be (i):$%
M_{I}=R\left( r\right) \oplus R\left( -r\right) X(x^{\prime })\widetilde{M}%
X(x^{\prime })R\left( r\right) \oplus R\left( -r\right) $ and (ii) $%
M_{II}=\Theta \left( -\theta \right) \oplus \Theta \left( -\theta \right)
X(x^{\prime })\widetilde{M}X(x^{\prime })\Theta \left( \theta \right) \oplus
\Theta \left( \theta \right) $. The EMs so generated are usually not in the
standard form, but this does not matter, after local squeezings $X(x)$ and $%
Y(y)$ being applied, the six parameter form EMs which are the most general
of $q-p$ decorrelation $M_{\sigma }$ will be generated. After minimization
of the $Tr\alpha _{\rho }M_{\sigma }$ with respect to $y$ and $x$ as
described in the former section and by using of the border state condition,
the relative entropy will become a function of $3$ variables which are $%
\gamma _{A},\gamma _{B},r$ for type (i) states and $\gamma _{A},\gamma
_{B},\theta $ for type (ii) states.

The type (iii) CM\ with $\frac{a}{b}+\frac{b}{a}=\frac{c_{1}}{c_{2}}+\frac{%
c_{2}}{c_{1}}$ and type (iv) CM\ with $a=b$ can be converted to their EMs by
solving the symplectic transformation matrix $S$ directly as in section 1.
The EM of type (iv) state is already given in section 1, with the border
state condition of $(m_{\sigma }-k_{\sigma q})(m_{\sigma }-k_{\sigma p})=1$,
the border EM will be in a form with only two variables $\gamma _{A},\gamma
_{B}$ through

\begin{equation}
s_{1}=\left( \frac{4\gamma _{A}^{2}\left( 4\gamma _{B}^{2}+1\right) }{%
4\gamma _{A}^{2}+1}\right) ^{\frac{1}{4}},\text{ \qquad }s_{2}=\left( \frac{%
4\gamma _{B}^{2}+1}{4\gamma _{B}^{2}\left( 4\gamma _{A}^{2}+1\right) }%
\right) ^{\frac{1}{4}}.
\end{equation}
The type (iii) CMs\ are of two kinds: $\frac{a}{b}=\frac{c_{1}}{c_{2}}$ and $%
\frac{a}{b}=\frac{c_{2}}{c_{1}}$. The $S_{q}$ matrices will be
\begin{equation*}
\left[
\begin{array}{ll}
\left( 1+\frac{\delta }{\gamma _{A}^{2}}\right) ^{\frac{1}{4}}, & 0 \\
\left( \frac{\delta ^{2}}{\gamma _{A}^{2}\left( \gamma _{B}^{2}+\delta
\right) }\right) ^{\frac{1}{4}}, & \left( \frac{\gamma _{B}^{2}}{\gamma
_{B}^{2}+\delta }\right) ^{\frac{1}{4}}
\end{array}
\right] ,\text{ and \quad }\left[
\begin{array}{ll}
\left( \frac{\gamma _{A}^{2}}{\gamma _{A}^{2}+\delta }\right) ^{\frac{1}{4}}
& \left( \frac{\delta ^{2}}{\gamma _{B}^{2}\left( \gamma _{A}^{2}+\delta
\right) }\right) ^{\frac{1}{4}} \\
0, & \left( 1+\frac{\delta }{\gamma _{B}^{2}}\right) ^{\frac{1}{4}}
\end{array}
\right]
\end{equation*}
for the two kinds after applying the border state condition for each, where $%
\delta =(\gamma _{A}^{2}-\frac{1}{4})(\gamma _{B}^{2}-\frac{1}{4})$.

The numerical results of all four type states indicate that GREE\ of the
state will be achieved by the border state of the same type. I calculate the
minimization of GRE of a given state with respect to all four type border
states. Some of the results of type (i) and type (ii) states are displayed
in Fig. (1) and Fig. (2). For type (iii) and type (iv) states, the border
state which achieves the GREE is also type (iii) state or type (iv) state
respectively. Moreover, it is worth noting that the state realized GREE has
the value $\left( \frac{a}{b}+\frac{b}{a}\right) /\left( \frac{c_{1}}{c_{2}}+%
\frac{c_{2}}{c_{1}}\right) $ which is very close to but not more than that
of the original state $\rho $. The numerical results are displayed in Fig.
(3). This adds evidence to the former numerical conclusion on types.
\begin{figure}[tbp]
\includegraphics[height=3in]{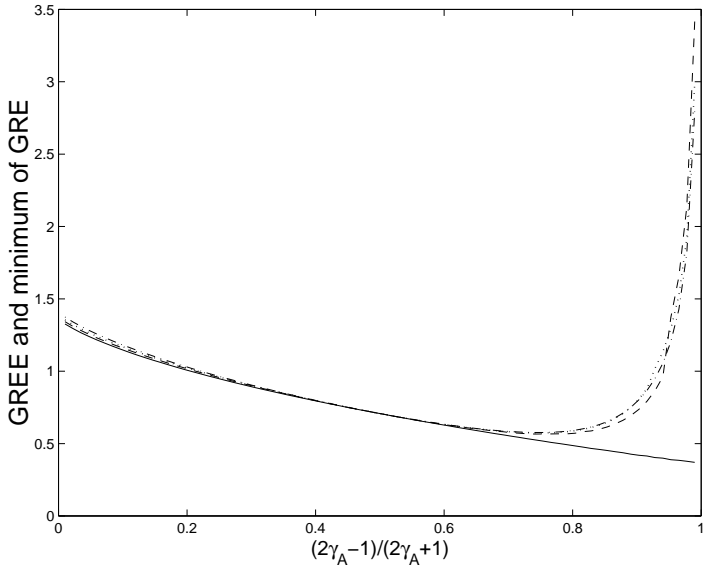}
\caption{For type (i) gaussian state with $%
sinh(2r)=sinh(2r_{border})+5,x=1.1 $, $(2\protect\gamma _{B}-1)/(2\protect%
\gamma _{B}+1)=0.5$. Solid line is for the searching result of type(i)
border states which achieve GREE, dash for type(ii) state reaching the
minimum of GRE within the type, dotted for type (iii),and dotted dash for
type (iv).}
\end{figure}

\begin{figure}[tbp]
\includegraphics[height=3in]{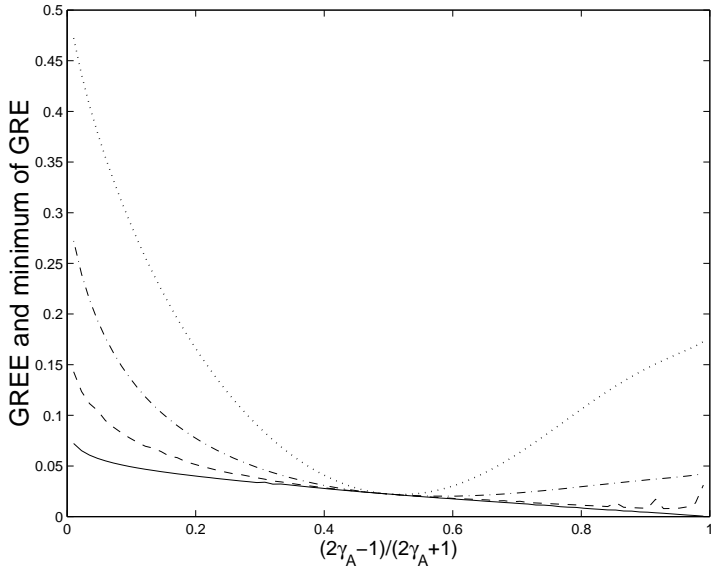}
\caption{For type (ii) gaussian state with $sinh(2\protect\theta%
)=0.5,x=x_{border}+1.5, (2\protect\gamma_{B}-1)/(2\protect\gamma_{B}+1)=0.5$%
. Solid line is for searching result of type(ii) border states which achieve
GREE, dash for type(i) state reaching the minimum of GRE within the type,
dotted for type (iii),and dotted dash for type (iv).}
\end{figure}

\begin{figure}[tbp]
\includegraphics[height=3in]{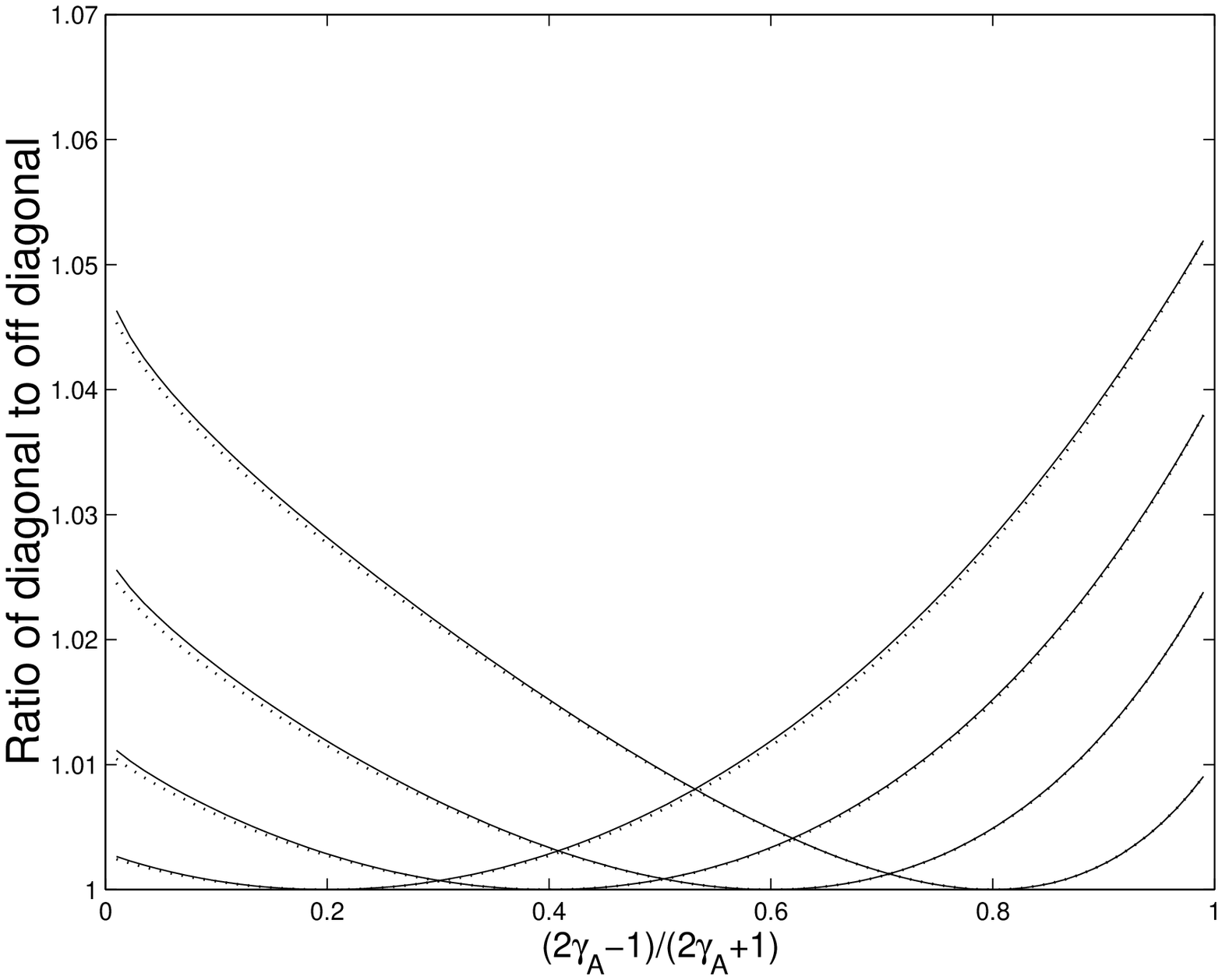}
\caption{Solid lines are for original gaussian states whose $%
sinh(2r)=sinh(2r_{border})+5,x=1.1$, dotted for the border states which
achieve GREE, In the left from up to down are for $(2\protect\gamma
_{B}-1)/(2\protect\gamma _{B}+1)=0.2,0.4,0.6,0.8$ respectively}
\end{figure}
One the other hand, for symmetric Gaussian state (the type (iii) state),
Based on the numerical observation, the conclusion can be drawn to be: for
each type of state, GREE\ will be achieved by the same type of state.

For type (iii) state $\rho _{S}$ (symmetric gaussian state) characterized by
CM\ of Eq. (\ref{wavelet}) and further by $m,k_{q},k_{p}$, the conclusion
that GREE is achieved by symmetric Gaussian state can be strictly
proven.Here I will give the main idea of proving. The detail will appear
elsewhere. Let us start with the 6 parameters $M_{\sigma }$ as was done at
the beginning of this section. This $6$ parameter EM can be produced as well
as reduced to $\widetilde{M}_{\sigma }=diag\{\widetilde{M}_{A},\widetilde{M}%
_{B},$ $\widetilde{M}_{A},\widetilde{M}_{B}\}$ by symplectic transformation $%
S,$ $S^{T}M_{\sigma }S=\widetilde{M}_{\sigma },$ Now the $S$
matrix has $4$ parameters which are independent of the symplectic
eigenvalues. The $S$ matrix takes the form of $S=S_{q}\oplus
(S_{q}^{T})^{-1}$ \ It is known that symplectic transformation can
always be dissolved to a rotation and squeezing then a successive
rotation, that is $S=R_{2}DR_{1}$. with $R_{1}$ and $R_{2}$ are
rotations and $D=diag\{d_{A},d_{B},1/d_{A},1/d_{B}\}$ is the
squeezing operation. Now for the $q-p$ decorrelation form of $S,$
$S=S_{q}\oplus
(S_{q}^{T})^{-1},$ $R_{1}$ and $R_{2}$ will be in their simple form of $%
R_{1}=\Theta \left( \theta \right) \oplus \Theta \left( \theta \right) ,$
and $R_{2}=\Theta \left( \phi \right) \oplus \Theta \left( \phi \right) ,$
By minimizing the relative entropy with respect to $\theta $ and $\phi $
under the restriction of $\sigma $ being a border state, one gets

\begin{equation}
\frac{\partial }{\partial \theta }\left( \frac{1}{2}Tr\alpha _{\rho
}M_{\sigma }+\lambda c_{1}c_{2}\right) =\frac{\partial }{\partial \phi }%
\left( \frac{1}{2}Tr\alpha _{\rho }M_{\sigma }+\lambda c_{1}c_{2}\right) =0,
\end{equation}
Where $\lambda $ is the multiplier. In general, the solution is quite
complicate and involved with the other four parameters $d_{A},d_{B},%
\widetilde{M}_{A},\widetilde{M}_{B}$. But for the symmetric Gaussian state $%
\rho _{S}$, the solutions are quite simple. They are (i) $\sin \theta =0,$ $%
\tan \phi =\pm 1;$(ii) $\cos \theta =0,$ $\tan \phi =\pm 1.$ And
the other four parameters
$d_{A},d_{B},\widetilde{M}_{A},\widetilde{M}_{B}$ are not
involved.\ The state $\sigma $ then will be the symmetric Gaussian
state.

The GREE will be
\begin{eqnarray}
E_{GR}\left( \rho _{S}\right)  &=&Tr\rho _{S}\log \rho _{S}+\min_{\widetilde{%
M}_{A},\widetilde{M}_{B}}\{-\sum_{j=A,B}\log (2\sinh \frac{\widetilde{M}_{j}%
}{2})  \label{wavelet2} \\
&&+\frac{1}{2}[(m+k_{q})(m-k_{p})\widetilde{M}_{A}^{2}+(m-k_{q})(m+k_{p})%
\widetilde{M}_{B}^{2}  \notag \\
&&+(m-k_{q})(m-k_{p})\widetilde{M}_{A}\widetilde{M}_{B}\coth \frac{%
\widetilde{M}_{A}}{2}\coth \frac{\widetilde{M}_{B}}{2}  \notag \\
&&+(m+k_{q})(m+k_{p})\widetilde{M}_{A}\widetilde{M}_{B}\tanh \frac{%
\widetilde{M}_{A}}{2}\tanh \frac{\widetilde{M}_{B}}{2}]^{\frac{1}{2}},
\notag
\end{eqnarray}
where $\widetilde{M}_{j}=\log \frac{2\gamma _{j}+1}{2\gamma _{j}-1},$ and $%
\gamma _{j}$ are symplectic eigenvalues of border state. Here it is easy to
prove that no further $X(x)$ operation is needed when the border state is
prepared in its EM\ with the form of Eq. (\ref{wavelet1}) (but with
different parameters), that is to say $x=1$.

If $k_{p}=k_{q}$, the state will be two mode squeezed thermal state\cite
{Chen} $\rho _{ST}$. Then Eq. (\ref{wavelet2}) is symmetric for $\widetilde{M%
}_{A},$ $\widetilde{M}_{B}$. Clearly the minimum will be achieved at $%
\widetilde{M}_{A}=$ $\widetilde{M}_{B}.$ So that
\begin{eqnarray}
E_{GR}\left( \rho _{ST}\right) &=&Tr\rho _{ST}\log \rho _{ST}+\min_{%
\widetilde{M}_{A}}\{-2\log (2\sinh \frac{\widetilde{M}_{A}}{2}) \\
&&+\frac{1}{2}\widetilde{M}_{A}[(m-k_{q})\coth \frac{\widetilde{M}_{A}}{2}%
+(m+k_{q})\tanh \frac{\widetilde{M}_{A}}{2}].  \notag
\end{eqnarray}

\section{Conclusions and Discussions}

Gaussian relative entropy of entanglement is an entanglement measure in its
own right. The relative entropy between two gaussian states was expressed as
correlation matrix of the first state and matrix in the exponential density
operator of the second state. The mutual transform of $q-p$ fashion CM\ and
EM was derived with symplectic transformation for $q-p$ decorrelation state.
A most general transform of CM\ to EM and vice versa was given through
commutation relation of the matrices and relation between the symplectic
eigenvalues of the matrices. I proved that gaussian relative entropy of
entanglement achieves when the separable gaussian state is at the border of
separable and inseparable sets. The displacement or first moments of the
second state can be ruled out as far as GREE\ is concerned. For GREE of $%
1\times 1$ gaussian state, the ten parameters EM\ of separable state which
minimizes the relative entropy was reduced to three variables EM. Where the
matrix was decomposed as local operations applied to a standard form of EM.
The three variables in EM\ were left for numerical calculation of the
minimization. To construct an EM\ more suitable for the calculation of GREE,
I\ classified the standard form CM of $1\times 1$ gaussian state into four
types according to some kind of ratio of diagonal to off diagonal for the
first three types and symmetry for the fourth. The numeric evidence on the
minimization of EMs strongly suggests that GREE for each type of gaussian
state will be realized by the state within the same type.

I strictly proved that GREE for symmetric Gaussian state is achieved by
symmetric gaussian state.It was given as the minimization of a function on
the two symplectic eigenvalues of EM. Although the minimization equations
are easily obtained, but they can not be solved analytically. Further more,
for a special kind of the symmetric state, the two mode squeezed thermal
state (TMST), the GREE will be a minimization of a function on one
parameter, the symplectic eigenvalue of TMST, and the state achieves the
GREE\ is a TMST state. I\ and my coworker had calculated \cite{Chen}the
minimization of relative entropy of TMST with respect to TMST as the upper
bound of relative entropy of entanglement, now it is proved that it is just
the GREE\ of TMST. Moreover, the upper bound for entanglement of formation
proposed in the same paper turns out to be the entanglement of formation
itself \cite{Giedke1}. So the comparison of the upper bound of EoF and RE of
TMST in our former paper \cite{Chen} is in fact the comparison of EoF and
GREE. And we had also provided coherent information and other entanglement
measure such as logarithmic negativity in that comparison.

I have given the method to calculate GREE for general state and the detail
calculation of GREE for $1\times 1$ gaussian state. It is expected that the
method developed in this paper will be applicable to the multi-mode
bipartite gaussian states and multipartite gaussian states.The definition of
\ GREE need not limit to gaussian state. For a non gaussian continuous
variable state, GREE can also be defined as the minimization of relative
entropy of the state with respect to all separable gaussian state. But there
is a deficiency that the relative entropy will never be zero. Never the
less, the calculation involves only the first and second moments of the
state, the necessary condition of separability on the CM of the non gaussian
state was also addressed \cite{Simon}, and the logarithmic of separable
gaussian state can be treated with EM.

Acknowledgment: Funding by the National Natural Science Foundation of China
(under Grant No. 10347119) is gratefully acknowledged.

\end{document}